\documentclass[runningheads]{svmult}

\usepackage{graphicx}  

\begin{document}
\title*{Chemical Composition of Planetary Nebulae: \newline
The Galaxy and the Magellanic Clouds*\newline\newline
\small * Planetary Nebulae beyond the Milky Way, ESO Astrophysics
Symposia, ed. J. R. Walsh, L. Stanghellini, Springer-Verlag, in press}
\toctitle{Chemical Composition of Planetary Nebulae: 
\protect\newline The Galaxy and the Magellanic Clouds}
%
%
\titlerunning{Chemical Composition of Planetary Nebulae}
%
\author{Walter~J.~Maciel
\and Roberto~D.~D.~Costa
\and Thais~E.~P.~Idiart}
\authorrunning{W. J. Maciel et al.}
%
%
\institute{IAG/USP, S\~ao Paulo SP, Brazil}

\maketitle              

\begin{abstract}
An analysis is made of the abundances of planetary nebulae (PN) in 
the Galaxy (MW), the Large Magellanic Cloud (LMC) and the Small Magellanic 
Cloud (SMC). The data has been gathered by the IAG/USP group in a 
homogeneous procedure, involving observational techniques, data 
acquisition and reduction. Emphasis is placed on distance-independent 
correlations, which are particularly important for PN in the galactic 
disk. It is shown that (i) our abundances are comparable to the 
accurate abundances of nebulae with ISO data, (ii) most abundance 
correlations hold for all three systems, taking into account the 
metallicity differences between the Galaxy and the Magellanic Clouds, 
and (iii) there is a good general agreement of our data with the 
predictions of recent theoretical models for PN progenitors.
\end{abstract}

\section{Introduction}
Planetary nebulae (PN) are especially important for studies
of the chemical evolution of the Galaxy and the Magellanic Clouds.
They are the offspring of stars in a mass range of roughly 0.8 to 8 
$M_\odot$, so that a relatively wide distribution of ages and 
population types can in principle be observed. 

Chemical abundances of PN are derived for several elements,
so that empirical distance-independent correlations can be 
determined with a reasonable degree of accuracy. These 
correlations can then be compared with the predictions of 
the nucleosynthetical processes and chemical evolution models
of intermediate mass stars (see for example \cite{marigo}, 
\cite{forestini}, \cite{groenewegen}) and act as constraints to 
these models.

In particular, it is known that there is a general metallicity 
difference between the Galaxy and the Magellanic Clouds, and between 
the LMC and SMC themselves. On  the other hand, it is not clear 
whether the chemical evolution has proceeded differently in the
Clouds as compared to the Galaxy, or the evolutionary steps 
were essentially the same, albeit starting from a more 
metal-poor gas. Some information in this respect can be obtained
by the analysis of the abundances of a relatively large sample 
of PN in these systems.

The total number of PN in the Clouds with accurate abundances is 
relatively small, as compared with the population of PN in 
the galactic disk and bulge (see for example \cite{stasinska}  
and \cite{leisy}). Since the early nineties, 
the IAG/USP group launched a long term project to derive accurate 
abundances for PN in the Galaxy and the Magellanic Clouds, based 
on detailed spectroscopic observations secured both in Brazil and 
in Chile (see \cite{costa2000} for some references). In this paper, 
we will present a brief account of this work.

\section{The data}
In a series of papers by the IAG/USP group, the chemical compositions 
of over two hundred planetary nebulae have been derived, including
objects in four different regions: (i) the galactic disk, 
(ii) the galactic bulge, (iii) the LMC and (iv) the SMC. 
The main references are \cite{costa2004}, \cite{ae2004},
\cite{ae2001}, \cite{costa2000}, \cite{fc2000}, \cite{costa1996},  
 \cite{fp1993a}, \cite{fp1993b}, \cite{fp1992}, and 
\cite{fp1991}. All observations have been obtained using either the 
1.6 m LNA telescope at Bras\'opolis, Brazil, or the 1.52 m ESO 
telescope at La Silla, Chile. At ESO a Boller and Chivens cassegrain 
spectrograph was used with a CCD detector at a reciprocal dispersion 
of about 2.5 \AA/pixel. At the LNA the dispersion was of about 
4.4 \AA/pixel. The reductions have been made using the IRAF package. 
Basically, the electron densities have been determinied from the [SII]
6716/6731\AA\ lines, while for the electron temperatures we have used the
[OIII] 4363/5007\AA\ and [NII] 5754/6584\AA\ lines. The observational techniques 
employed, data acquisition and reduction procedures have all been similar, 
which warrants the high degree of homogeneity desirable in the analysis 
of the chemical composition of planetary nebulae. 

\section{Distance-independent correlations}
\subsubsection*{Planetary nebulae and HII regions.}
Comparing the average abundances of PN with the corresponding values for 
HII~regions, we find that the abundances of the elements S, Ne, Ar, and 
O are similar both in the PN and HII regions of each system. Average
values for oxygen in PN are $\epsilon({\rm O}) = \log$ O/H + 12
$\simeq$ 8.65, 8.40 and 8.16 dex for the MW, LMC and SMC, respectively. 
The differences for O, S, Ar and Ne are relatively small, generally under 
0.2 dex, similar or smaller than the average abundance 
uncertainties. The main differences appear for 
helium and especially nitrogen, which are enriched in PN, and are dredged up 
in the progenitor stars. The differences amount to about 40\% for He/H and 
0.5 to 0.7 dex for N/H. 

\subsubsection*{Correlations involving S, Ar, Ne and O.}
The elements S, Ar and Ne are not expected to be produced by the PN progenitor 
stars, as they are manufactured in the late evolutionary stages 
of massive stars. Therefore, S, Ar and Ne abundances as measured in 
planetary nebulae should reflect the interstellar composition at the time 
the progenitor stars were formed. Regarding the oxygen abundances, there 
are some evidences of a reduction in the O/H ratio in some PN due to ON cycling 
in their progenitor stars, so that there may be some small variations in 
the O/H abundances in the nebular gas relative to the pristine abundances. 

The variation of the ratios S/H, Ar/H and Ne/H with O/H show a good positive 
correlation for all systems, with similar slopes. Our sample is not
complete, so that the average abundances do not necessarily reflect
the metallicities of the galaxies considered. However, the metallicity
differences can be observed judging from the observed metallicity
range. The galactic nebulae extend to higher metallicities 
(up to $\epsilon({\rm O}) \simeq$ 9.2), while the LMC objects 
reach $\epsilon({\rm O})\simeq$ 8.8, and the SMC extends to even lower 
metallicities, namely $\epsilon({\rm O}) \simeq$ 7.0. 
Taking into account the galactic and MC nebulae by 
Stasi\'nska et al. \cite{stasinska}, these results are confirmed and
extended, especially at lower metallicities, $\epsilon({\rm O})<$ 8.0, for
which their sample is richer than ours.

\subsubsection*{Correlations involving N.}
An anticorrelation between N/O and O/H has been discussed 
by a number of people (see for example \cite{costa2000}, 
\cite{stasinska} and Perinotto et al., this conference)
and may be an evidence for the conversion of O into N in the PN progenitor stars. 
This scenario is basically supported by our data. The galactic disk nebulae 
show a mild correlation, taking into account the majority of objects 
with $\epsilon({\rm O}) >$ 8.0. For the Magellanic Clouds, the anticorrelation 
appears better defined, particularly for the SMC. In this case,
our sample fits nicely the previous sample \cite{stasinska}, 
especially at higher O/H ratios, that is, $\epsilon({\rm O}) >$ 8.0. 

Assuming that the nitrogen abundance is directly proportional to the 
oxygen abundance, as is the case for S, Ar and Ne, then for lower O/H 
ratios the N/O ratio is expected to be essentially constant. Since this ratio 
increases in average for lower O/H values, we have to conclude that the 
nitrogen abundance is enhanced, possibly at the expenses of oxygen. 
Since the N/O $\times$ O/H anticorrelation is better defined for the
Magellanic Clouds, it seems that the lower metallicities of these
galaxies favour the ON-cycling as observed in planetary nebulae.

A well defined N/O $\times$ N/H correlation is shown in figure~1
for galactic (crosses), LMC (empty circles) and SMC (filled circles)
nebulae. Average error bars are provided at the lower right corner.
The dashed line shows the theoretical predictions of the
final model for the LMC by Groenewegen and de Jong \cite{groenewegen}. 
The inflection point of the  line corresponds to initial 
stellar masses of about 2 $M_\odot$, lower masses being associated with the 
left side of the plot and higher masses to the right side. As pointed out in
\cite{groenewegen}, the location of PN in this diagram
is a good indicator of main sequence mass of the progenitor star.
We can see that our SMC sample includes basically low mass progenitors. 

\begin{figure}
\begin{center}
\includegraphics[angle = -90, width= .85\textwidth]{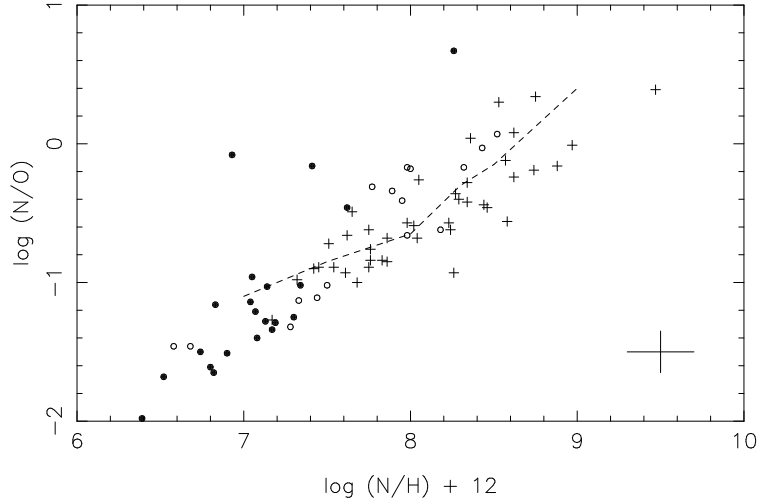}
\end{center}
\caption[]{Nitrogen abundances of PN in the Galaxy (crosses),
   LMC (empty circles) and SMC (filled circles). The dashed
   line shows predictions of the final model by Groenewegen 
   and de Jong \cite{groenewegen}. Average error bars are
   shown at the lower right corner.}
\label{fig1}
\end{figure}

\subsubsection*{Correlations involving He.}
Another interesting correlation is a plot of N/O as a function of 
He/H (figure~2). For the Galaxy (crosses), a good 
correlation is observed. The inclusion of the LMC and SMC objects generally 
supports this conclusion, although the scattering increases, probably 
due to the different metallicities of the galaxies. The main 
differences occur for the SMC, for which the He/H ratio has 
approximately the same range of the LMC and even the Galaxy, but N/O 
is generally lower, at least in our sample. Since the O/H ratio is 
also smaller in the SMC, due to the metallicity difference, we 
apparently observe a smaller nitrogen enhancement in the SMC, so 
that the progenitor star masses are somewhat lower in this galaxy,
as also suggested by Fig.~1. In other words, type~I PN are probably less 
frequent in the SMC, in agreement with the conclusion by Stanghellini 
et al. \cite{stanghellini}, based on HST data, that bipolar PN are 
much rarer in the SMC than in the LMC or in the Galaxy (see also
Shaw and Villaver, this conference). It is argued 
that the lower metallicity environment of the SMC does not favour 
the formation of the higher mass progenitors that give rise to Type I 
bipolar (or asymmetric) PN.

\begin{figure}
\begin{center}
\includegraphics[angle = -90, width= .85\textwidth]{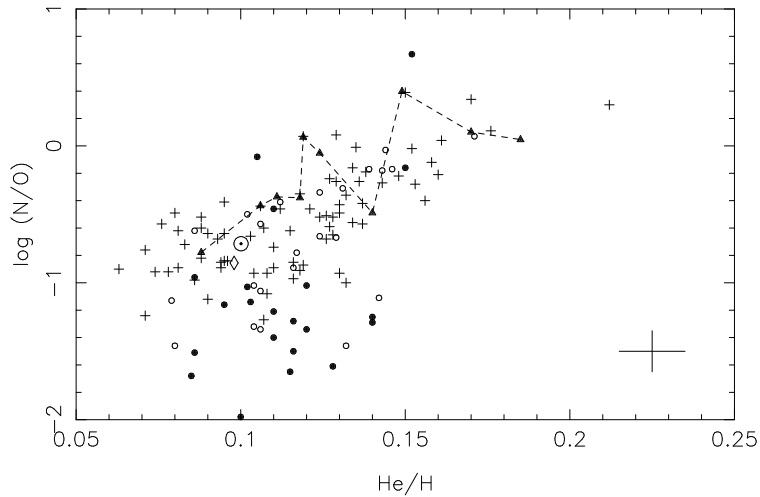}
\end{center}
\caption[]{The N/O $\times$ He/H correlation for PN. Symbols are
   as in Fig. 1. Also included are the Sun, Orion (losangle) and 
   galactic PN with ISO data from Marigo et al. \cite{marigo} 
   (triangles connected by dashed lines).}
\label{fig2}
\end{figure}

Recently, Marigo et al. \cite{marigo} have presented abundances
of galactic PN for which ISO and IUE spectra are available, 
so that their abundances are presumably well determined, as all important 
ionization stages have been considered, making it practically unnecessary 
to use ionization correction factors. Also, infrared data are less dependent 
on the electron temperatures and their intrinsic uncertainties.
These objects are included in figure~2 (triangles connected by dashed lines), 
and show a good agreement with our data, at least for the Milky Way. 
Another comparison of our galactic PN with nebulae having ISO 
abundances can be seen in figure~3, where we plot our PN
along with nebulae having ISO abundances from Pottasch et al. \cite{pottasch}
(triangles connected by dashed lines). Also shown in the figure are the 
Sun, Orion (losangle) and predictions of theoretical models by Marigo et al. 
\cite{marigo} (continuous line), which consist of synthetic evolutionary 
models for the thermally pulsing asymptotic giant branch stars.

\begin{figure}
\begin{center}
\includegraphics[angle = -90, width= .85\textwidth]{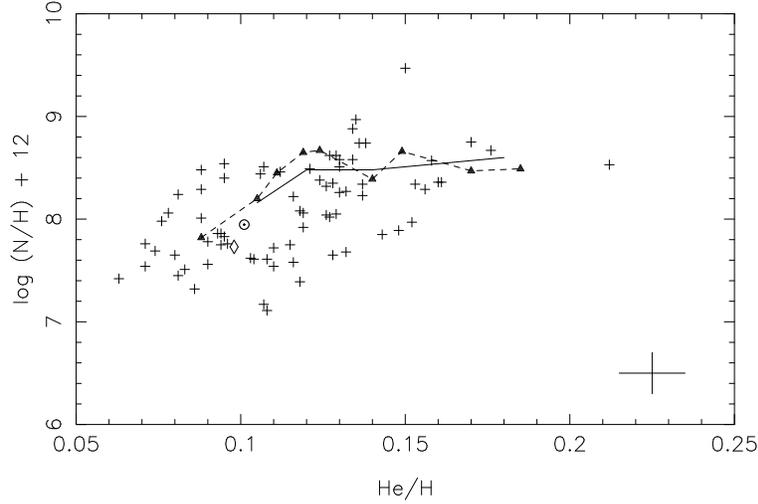}
\end{center}
\caption[]{The N/H $\times$ He/H correlation for galactic PN. 
   Also included are the Sun, Orion (losangle),  PN 
   with ISO data from Pottasch et al. \cite{pottasch} (triangles
   connected by dashed lines) and predictions of theoretical models 
   by Marigo et al. \cite{marigo} (continuous line).}
\label{fig3}
\end{figure}

\section{Final remarks}
Recently, Costa et al. \cite{costa2004} have analyzed a sample of PN 
located in the direction of the galactic anticenter. It was shown that the 
observed radial O/H gradient tends to flatten out for large galactocentric 
distances, roughly $R > 10$ kpc, in agreement with previous results by 
Maciel and Quireza \cite{mq1999}. Therefore, it is interesting to 
investigate whether anticenter PN -- which are in principle more 
metal poor than their inner counterparts -- have abundances closer to  
metal poor systems, namely  the LMC and SMC. 

An analysis of our sample shows that all galactic nebulae, comprising 
inner and outer objects, are evenly distributed along the disk, so that 
there is apparently no association of the outer nebulae with the lower 
metallicity objects of the Magellanic Clouds. However, taking into account
the nebulae of the larger sample of Lago and Maciel (in preparation), which
includes objects from the literature at low galactocentric distances, 
we conclude that a tendency can be observed, 
in the sense that the higher metallicity objects are preferentially located 
closer to the galactic center.

Finally, it is interesting to notice that the same 
correlations found for PN in the galactic disk and in the Magellanic Clouds
can also be observed in the objects of the galactic bulge, as recently
shown by Escudero et al. \cite{ae2004}.

\bigskip
{\it Acknowledgements.} 
We thank J. A. de Freitas Pacheco, M. M. M. Uchida, L. G. Lago and
A. V. Escudero for some helpful discussions. This work was partially 
supported by CNPq and FAPESP.

%

\end{document}